\documentclass[a4paper,11pt]{article}
\pdfoutput=1 

\usepackage{jcappub} 

\usepackage[T1]{fontenc} 
\def\mnras{MNRAS}
\def\apj{ApJ}
\def\apjl{ApJL}

\def\aap{A\&A}

\def\nat{Nature}

\title{\boldmath Gravitational wave emission by the high braking index pulsar PSR J1640-4631}

\author{Jos\'e C. N. de Araujo}
\author{Jaziel G. Coelho,}
\author{Cesar A. Costa}

\affiliation{Divis\~ao de Astrof\'isica, Instituto Nacional de Pesquisas Espaciais, Avenida dos Astronautas 1758, S\~ao Jos\'e dos Campos, 12227--010 SP, Brazil}

\emailAdd{jcarlos.dearaujo@inpe.br}
\emailAdd{jaziel.coelho@inpe.br}
\emailAdd{cesar.costa@inpe.br}

\abstract{Recently, a braking index for the pulsar PSR J1640-4631 has been measured. With a braking index of $n = 3.15 \pm 0.03$, this pulsar has the highest braking index ever measured. As it is well known, a pure magnetic dipole brake yields $n = 3$, whereas a pure gravitational wave (GW) brake yields $n = 5$. Therefore, each of these mechanisms alone can not account for the braking index found for PSR J1640-4631. Here we consider in detail that such a braking index could be accounted for if the spindown model combines magnetic dipole and GW brakes. Then, we briefly discuss the detectability of this pulsar by  aLIGO and the planned Einstein Telescope. In particular, we show that the amplitude of the GW that comes from our model is around a factor four lower than the amplitude modeled exclusively by GW energy loss. Another interesting outcome of our modeling is that it is possible to obtain the ellipticity from the braking index and other pulsar parameters.}

\begin{document}
\maketitle
\flushbottom

\section{Introduction}
\label{int}
One hundred years after being predicted and decades of experimental efforts, the gravitational waves (GWs) have been finally detected \cite{2016PhRvL.116f1102A}. A signal was observed by the LIGO detectors and comes from the coalescence of a binary black hole system. Besides the binary systems composed of compact stars, there are many other sources of GWs, among them the pulsars (spinning neutron stars) that could well be detected in the near future.

The so called braking index, which is closely related to the pulsar's spindown, can provide information about the energy loss of these objects. Until very recently, only eight of $\sim 2400$ known pulsars have measured braking indices with values ranging from $0.9\pm0.2$ to $2.839\pm0.001$~
\citep[see e.g.,][]{1993MNRAS.265.1003L,1996Natur.381..497L,2007ApSS.308..317L,2011MNRAS.411.1917W,2011ApJ...741L..13E,2012MNRAS.424.2213R, 2015ApJ...810...67A}. The magnetic dipole assumption predicts a braking index $n=3$. Several interpretations of the observed braking indices have been put forward, like the ones that propose either accretion of fall-back material via a circumstellar disk \citep{2016MNRAS.455L..87C}, relativistic particle winds \citep{2001ApJ...561L..85X,2003A&A...409..641W}, or modified canonical models to explain the observed braking index ranges \citep[see e.g.,][]{1997ApJ...488..409A,2012ApJ...755...54M,2016arXiv160301487E}, and references therein for further models). Alternatively, it has been proposed that the so-called quantum vacuum friction (QVF) effect in pulsars can explain several aspects of their phenomenology~\citep{2008EL.....8269002D, 2012EL.....9849001D, CoelhoQVF}.
However, no model has been developed yet that explains satisfactory all measured braking indices, nor any of the existing ones has been totally ruled out by current data. Therefore, energy loss mechanisms for pulsars are still under debate.

Recently, \citet{2016ApJ...819L..16A} showed that the PSR J1640-4631 is the first to have a braking index greater than the canonical value (three), $n=3.15\pm 0.03$. PSR J1640-4631 has a spin period of $P=206$~ms and a spindown rate of $\dot P=9.758(44)\times 10^{-13}$~s/s, yielding a spindown power $\dot E_{\rm rot}=4.4\times 10^{36}$~erg/s, and inferred dipole magnetic field $B_0=1.4\times10^{13}$~G. This source was discovered using X-ray timing observations from NuStar and a measured distance of $\sim 12$~kpc~\citep[see][]{2014ApJ...788..155G}.
 
It is worth noticing that LIGO and VIRGO have released their results on the search for continuous GWs putting forward upper limits for the sources ellipticities \citep{2015arXiv151003621T,2015ApJ...813...39A,PhysRevD.90.062010}. Their results show that the typical ellipticity would be $\varepsilon < 2 \times 10^{-5} $. Moreover, as already mentioned, aLIGO (advanced LIGO) has just completed its first observational run (O1) and observed the first GW transient \citep{2016PhRvL.116f1102A}. Now $\sim 4$ months are being analyzed and new results for the search for continuous signals might be released soon, according to one of the authors of this article, Cesar A Costa, who is also member of the LIGO Collaboration.

In this paper, we are concerned about PSR J1640-4631 in two different ways: {\it{i}}) the electromagnetic emission of a neutron star derived from its rotational kinetic energy, and its spindown is usually measured in terms of a braking index, $n$, which is dependent on the magnetic field configuration; {\it{ii}}) on the other hand, pulsars can also spindown through gravitational emission associated to asymmetric deformations. The observed rotational energy loss provides a huge reservoir of energy, along with magnetic dipole radiation some fraction of this reservoir is dissipated through GW emission~\citep[see e.g.,][]{1969ApJ...157.1395O,1969ApJ...158L..71F}.

Also, the recently braking index $n=3.15$ measured for the rotationally powered pulsar PSR J1640-4631 reignites the question about the fundamental energy loss mechanisms of the pulsars. Our interest in this work is to revisit the issue of the gravitational and electromagnetic contributions in the context of pulsars with putative $n>3$. 
This paper is organized as follows. In the next section we revisit the fundamental energy loss mechanisms for pulsars. We also derive its associated energy loss focusing mainly on the energy balance and model self-consistency when both gravitational and classic dipole radiation are responsible for the PSR J1640-4631 spindown. In Section \ref{sec:summary}, we summarize the main conclusions and remarks. We work here with Gaussian units.

\section{The energy balance of PSR J1640-4631}
As already mentioned, we consider that the main energy loss sources of PSR J1640-4631 (or any other putative pulsar with  $n > 3$) are magnetic dipole brake and GW emission.

Recall that if the pulsar magnetic dipole moment is misaligned with its spin axis by an angle $\phi$, the energy emitted per second by a rotating magnetic dipole reads \citep[see e.g.,][]{2001thas.book.....P, 1975ctf..book.....L},
\begin{equation}
\dot{E}_{\rm d}= \frac{16\pi^4}{3}\frac{B_0^2 R^6\sin^2\phi}{P^4c^3},  \label{Ed}
\end{equation}
where $B_0$ is the mean surface magnetic field of a star of radius $R$ and rotational period $P$.

Spinning neutron stars which possess asymmetric deformations emit GWs. More precisely, a spheriodal body with moment of inertia, $I$, and equatorial ellipticity, $\epsilon$, emits GWs. In this case, the energy loss via GW emission reads~\citep[see e.g.,][]{1983bhwd.book.....S}
\begin{equation}
\dot{E}_{\rm GW} = \frac{2048\pi^6}{5}\frac{G}{c^5}\frac{I^2\epsilon^2}{P^6}. \label{EGW}
\end{equation}
An absolute upper limit on the GW strain from isolated pulsars, known as the spindown limit, can be calculated assuming that the observed loss of rotational energy ($\dot E_{\rm rot} = I \Omega \dot\Omega$) is all going into gravitational radiation, i.e. $\dot{E}_{\rm GW}$~\citep[see, e.g.,][]{2014ApJ...785..119A}.
Instead, we consider in this paper that the total energy emitted by the star is provided by its rotational counterpart, $E_{\rm rot}=I\Omega^2_{\rm rot}/2$, and any change on it would be attributed to both $\dot{E}_{\rm d}$ and $\dot{E}_{\rm GW}$, namely 
\begin{equation}
\dot{E}_{\rm rot}\equiv \dot{E}_{\rm GW} +\dot{E}_{\rm d}.\label{Erotdef}
\end{equation}
Since $\Omega_{\rm rot} = 2\pi/P$, it follows immediately that
\begin{equation}
\dot{\Omega}_{\rm rot} = \frac{32}{5} \frac{G}{c^5}I\epsilon^2\Omega^5_{\rm rot}+\frac{1}{3}\frac{B_0^2R^6\sin^2\phi}{Ic^3}\Omega^3_{\rm rot}.\label{decel}
\end{equation}
This equation can be interpreted as follows: the term on the left side stands for the resulting deceleration (spindown) due to the emission of GWs and the magnetic brakes, the first and second terms on the right side represent the independent contributions of each one of these processes. The above equation can be conveniently rewritten as follows
\begin{equation}
\dot{\Omega}_{\rm rot} = \dot{\Omega}_{\rm GW} + \dot{\Omega}_{\rm d}. 
\end{equation}
It is useful to define the fraction of deceleration $(\eta)$ related to the GW emission, namely
\begin{equation}
\eta = \frac{\dot{\Omega}_{\rm GW}}{\dot{\Omega}_{\rm rot}}.
\end{equation}
Substituting the appropriate quantities one has
\begin{equation}
\eta = \frac{1}{1+\frac{5}{384}\frac{c^2B_0^2R^6\sin^2\phi}{G\pi^2 I^2\epsilon^2}P^2}.\label{eta}
\end{equation}
Notice that the above definition implies that $\dot{\Omega}_{\rm GW} =  \eta \dot{\Omega}_{\rm rot}$. Thus $\eta$ can also be interpreted as the fraction of the power lost by the pulsar in the form of GWs, or also the efficiency of generation of GWs.
It is appropriate to rewrite the equation for $\eta$ in terms of the braking index, that is given by
\begin{equation}
n = \frac{\Omega_{\rm rot}\,{\ddot\Omega}_{\rm rot}}{\dot{\Omega}^2_{\rm rot}}\label{n}.
\end{equation}

Before proceeding it is worth recalling that a pure magnetic brake, in which a dipole magnetic configuration is adopted, gives $n = 3$, whereas a pure GW brake gives $ n = 5$.
Therefore, neither a pure GW brake nor a pure magnetic dipole brake are not supported by the observations. On the other hand, a combination of both processes of energy loss considered in the present paper could account for the braking index of, for example, PSR J1640-4631 (or any other putative pulsar with $n > 3$).

Substituting equation \ref{decel} and its derivative in equation \ref{n} one has
\begin{equation}
n = 3 + \frac{2}{1+\frac{5}{384}\frac{c^2B_0^2R^6\sin^2\phi}{G\pi^2 I^2\epsilon^2}P^2}.\label{ns}
\end{equation}

This equation naturally leads to values of the brake indices $3 \leq n \leq 5$. Notice that combining equations \ref{eta} and \ref{ns}, one obtains
\begin{equation}
\eta = \frac{n-3}{2}.\label{etan}
\end{equation}
Therefore, $\eta$ is directly related to the observable quantity $n$. An immediate consequence thereof is that 
\begin{equation}
\dot{\Omega}_{\rm GW} = \eta  \dot{\Omega}_{\rm rot} = \frac{(n-3)}{2}\, \dot{\Omega}_{\rm rot}. 
\end{equation}
Since the angular velocity is directly related to $\dot{f}_{\rm rot}$, the above equation can be rewritten in the following form
\begin{equation}
\dot{\bar{f}}_{\rm rot} = \frac{(n-3)}{2}\,\dot{f}_{\rm rot}, \label{dbf}
\end{equation}
where we can interpret $\dot{\bar{f}}_{\rm rot}$ as the part of $\dot{f}_{\rm rot}$ that contributes to the generation of GWs.

Now, we consider how the amplitude of the GWs for pulsars with $n < 5$ can be calculated. Recall that one usually finds in the literature the following equation
\begin{equation}
h^2 = \frac{5}{2}\frac{G}{c^3}\frac{I}{r^2}\frac{\vert \dot{f}_{\rm rot}\vert}{f_{\rm rot}} \label{h5}
\end{equation}
\citep[see, e.g.,][]{2014ApJ...785..119A}, where the whole contribution to $\dot{f}_{\rm rot}$ comes from the GW emission, i.e., its implicitly assumed that $n =5$. This equation must be modified to take into account that $n < 5$. To do so the equation for the amplitude of the GW can be written in the following form
\begin{equation}
\bar{h}^2 = \frac{5}{2}\frac{G}{c^3}\frac{I}{r^2}\frac{\vert \dot{\bar{f}}_{\rm rot}\vert}{f_{\rm rot}} =   \frac{(n-3)}{2}\, h^2,\label{hn}
\end{equation}
where equation \ref{dbf} was substituted in the last equality.

From the above equation one has for PSR J1640-4631 that $\bar{h} \simeq 0.27 h$, i.e., almost a factor of four lower than the amplitude found when one assumes that the energy loss is completely given by GW emission.
Notice that in the present case, since $n=3.15$, $\eta =0.075$, which means that the GW luminosity would be $7.5 \%$ of the total power lost ($\dot{E}_{\rm rot}$). 

In addition, starting from
\begin{equation} 
h = \frac{16\pi^2G}{c^4} \frac{I\epsilon f_{\rm rot}^2}{r}
\end{equation}
\citep[see, e.g.,][]{1983bhwd.book.....S} and equation \ref{hn}, one readily 
obtains an equation for $\epsilon$ in terms of $n$, $P$, $\dot P$ (observable quantities) and $I$, namely
\begin{equation}
\epsilon = \sqrt{\frac{5}{1024\pi^4} \frac{c^5}{G}\frac{\dot{P}P^3}{I}(n-3)}.
\end{equation}
Notice that this model predicts that the ellipticity would be a factor of $\sqrt{(n-3)/2}$ smaller than that when one assumes that the energy loss is given only in terms of GW emission.

Assuming that $I \approx 10^{38} \, \rm{kg \, m^2}$ (fiducial) and substituting the values of $n$, $P$, $\dot P$ for PSR J1640-4631, we obtain $\epsilon \simeq 4.8\times 10^{-3}$. One may wonder if such a high ellipticity could be justifiable without considering exotic models. Whether it is not possible to explain such a figure appropriately, this could be an indication that other mechanisms, apart of GW and dipole magnetic brakes must necessarily be considered. As a consequence thereof it could well occur that $\eta \ll 0.075 $ implying that $\epsilon \ll 10^{-3}$, or vice-versa.

\begin{figure}[!htb]
\includegraphics[width=\linewidth,clip]{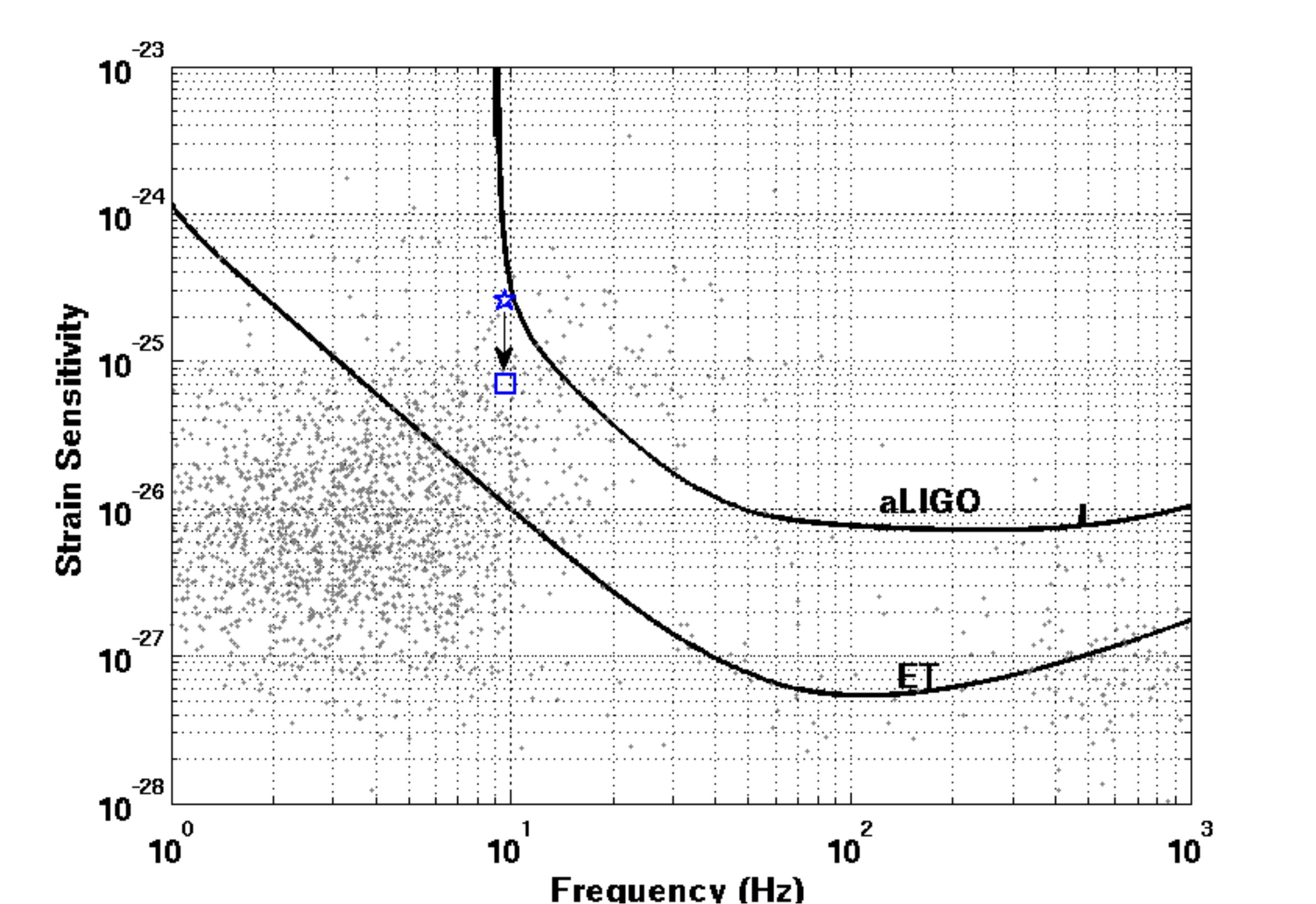} 
\caption{Strain sensitivities for aLIGO and ET for one year of integration time and the strain amplitudes for PSR J1640-4631 using equations \ref{h5} (star) and \ref{hn} (square). The cloud of dots represents 1880 pulsars with strain calculated by \ref{h5} \citep[found in  ][]{{ATNF}}.}\label{fig1}
\end{figure}

Anyway, it is interesting to see if either aLIGO or the planned Einstein Telescope (ET) could detect PSR J1640-4631, in the context here studied. In figure \ref{fig1} we show the strain for PSR J1640-4631 using equations \ref{h5} (star) and \ref{hn} (square) and the strain sensitivities curves for aLIGO and ET for one year of integration time \citep{2010CQGra..27a5003H, 2015CQGra..32g4001L}. This pulsar emits GWs at $f_{\rm GW} = 2 / P \simeq 9.7 \; \rm{Hz}$, where aLIGO is not sensitive enough to detect it even for one year of integration time. On the other hand, ET, for the same integration time, could well detect it. Notice that the cloud of dots represents the strain calculated by equations \ref{h5} for 1880 pulsars from ATNF Pulsar Catalog.

\section{Summary}\label{sec:summary}
In this paper we model the PSR J1640-4631 spindown by means of a combination of energy loss mechanisms which includes GW emission and magnetic dipole brake. We have shown that with this modeling it could be possible to account for this pulsar braking index. But, in this case it is mandatory to explain how it is possible that a pulsar have such a high ellipticity.

Concerning the detectability of PSR J1640-4631 via its putative gravitational emission, we conclude that aLIGO, even for one year integration time, would not observe it. On the other hand, since the planned ET is more sensitivity than aLIGO at such frequency, it would be able to detect PSR J1640-4631 with the appropriate integration time. Bearing in mind the high ellipticity implicit in this calculation.

An interesting question that deserves to be appropriately addressed has to do with the modeling of all other eight pulsars with  accurately measured braking indices. Due to their dynamic nature, pulsars should always present important temporal changes in quantities other than $P$, such as $B_0$ and $\phi$. Moreover, since the pulsars' ellipticities are very likely non null, the contribution of the GW brake needs necessarily to be considered. We argue that, no matter what are the other mechanisms considered in order to explain the measured braking index, the GW contribution must necessarily be taken into account.

Last, but not least, a model that takes into account, besides the GW and the magnetic dipole brakes, $B_0$ and $\phi$ dependent on time, could also provide a picture in which the braking index of PSR J1640--4631 could be explained without the need of such a high ellipticity. These issues are part of a study to appear elsewhere.

\section*{Acknowledgements}

J.C.N.A thanks FAPESP (2013/26258-4) and CNPq (308983/2013-0) for partial support. J.G.C. acknowledges the support of FAPESP (2013/15088-0 and 2013/26258-4). C.A.C. acknowledges CNPq (158428/2014-5) and PNPD-CAPES for financial support.


\end{document}